\documentstyle[a4,12pt]{article}

\renewcommand{\thefootnote}{\fnsymbol{footnote}}

\def\beq{\begin{equation}}
\def\eeq{\end{equation}}
\def\bea{\begin{eqnarray}}
\def\eea{\end{eqnarray}}
\def\nn{\nonumber}
\def\dfrac{\displaystyle\frac}

\def\btiny{\begin{tiny}}
\def\etiny{\end{tiny}}
\def\bsc{\begin{scriptsize}}
\def\esc{\end{scriptsize}}
\def\bfoot{\begin{footnotesize}}
\def\efoot{\end{footnotesize}}
\def\bsm{\begin{small}}
\def\esm{\end{small}}
\def\bno{\begin{normalsize}}
\def\eno{\end{normalsize}}
\def\bla{\begin{large}}
\def\ela{\end{large}}
\def\bLa{\begin{Large}}
\def\eLa{\end{Large}}
\def\bLA{\begin{LARGE}}
\def\eLA{\end{LARGE}}
\def\bhu{\begin{huge}}
\def\ehu{\end{huge}}
\def\bHu{\begin{Huge}}
\def\eHu{\end{Huge}}

\def\bCe{\begin{center}}
\def\eCe{\end{center}}
\def\bFR{\begin{flushright}}
\def\eFR{\end{flushright}}
\def\bFL{\begin{flushleft}}
\def\eFL{\end{flushleft}}

\def\eqref#1{eq.(\ref{eqn:#1})}
\def\eqlab#1{\label{eqn:#1}}
\def\tbref#1{table(\ref{table:#1})}
\def\tblab#1{\label{table:#1}}

\def\bmaT{\left(\begin{array}{ccc}}
\def\emaT{\end{array}\right)}
\def\bma{\left( \begin{array} }
\def\ema{\end{array} \right)}

\def\sb {\mbox{\footnotesize $\Box$}}
\def\sbb {{\mbox{\footnotesize $\Box$}}^\ast}
\def\ve {\varepsilon}

\def\pq  {P_{\rm Q}}
\def\pl  {P_{\ell}}
\def\pu  {\bar{P}_{\rm U}}
\def\pd  {\bar{P}_{\rm D}}
\def\pn  {\bar{P}_{\rm N}}
\def\pe  {\bar{P}_{\rm E}}

\def\phiu {\bar{\phi}_{\rm u}}
\def\phid {\bar{\phi}_{\rm d}}
\def\Phiu {\bar{\Phi}_{\rm u}}
\def\Phid {\bar{\Phi}_{\rm d}}

\def\Gu {G_{\rm u}}
\def\Gd {G_{\rm d}}
\begin{document}
\title{Yukawa Interaction \\ from a SUSY Composite Model}
\author{{N. Haba $^{1,2}$}\thanks{E-mail address : haba@phen.mie-u.ac.jp}{\quad}and{\quad}{N. Okamura$^{1}$}\thanks{E-mail address : okamura@eken.phys.nagoya-u.ac.jp}\\ \\{\it $^1$Department of Physics, Nagoya University,}\\{\it Nagoya, Japan, 464-01}\\ \\{\it $^2$Faculty  of Engineering,  Mie University,}\\{\it Mie, Japan, 514}}
\date{September, 1997}
\maketitle
\vspace{-11.5cm}
\begin{flushright}
DPNU-97-27\\
Sep., 1997
\end{flushright}
\vspace{9.0cm}
\begin{abstract}
We present a composite model
that is based on non-perturbative effects of
$N=1$ supersymmetric $SU(N_C)$ gauge theory 
with $N_f=N_C+1$ flavors.
In this model, 
we consider $N_C=7$, 
where all matter fields in the supersymmetric
standard model, that is,
quarks, leptons and
Higgs particles are bound states of
preons and anti-preons.
When $SU(7)_H$ hyper-color coupling 
becomes strong,
Yukawa couplings of
quarks and leptons 
are generated dynamically.
We show one generation model at first,
and next we show models of three generations.
\end{abstract}

\newpage

\section{Introduction}
\label{sec:1}
\setcounter{equation}{0}
\hspace*{\parindent}
The origin of Yukawa couplings
is one of the greatest mystery 
in the Standard Model (SM).
This situation is not changed
in the supersymmetric standard model (SSM).
A search for the origin of the Yukawa couplings
might show us the way to get at the origin of the masses
and the Kobayashi-Maskawa (KM) flavor mixing matrix 
{\cite {KM}}.

There has been an idea 
that quarks and leptons may be composite particles
and Yukawa terms are induced by gauge dynamics.
We revive this idea using supersymmetric QCD.
According to the recent progress of 
$ N=1 $ supersymmetric ${SU(N_C)}$ 
with ${N_f = N_C+1}$ gauge theories
{\cite{Seiberg,Reviews1}}
the non-perturbative effects
induce the superpotential
\beq
W_{\rm dyn}=\dfrac{1}{\Lambda^{2N_C-1}}\left(\bar{B}MB-\det{M}\right).
\eqlab{11}
\eeq
In this case, the moduli space is not changed by quantum corrections.
This superpotential might suggest the possibility that
matter fields are composite states
and 
Yukawa couplings emerge from the strong dynamics
when baryon (anti-baryon) states are regarded as
quarks and leptons
and
Higgs particles regard as meson states.

Recently,
many authors have considered various models
in which SSM particles are composed from 
more elementary particles and
Yukawa couplings are induced strong gauge dynamics \eqref{11}
{\cite{nelson,nelson2,luty,luty2,kitazawa,hayakawa}}
.

In ref.{\cite{nelson}},
right-handed down quarks ${\bar{D}_R}$
and $SU(2)_L$ doublet leptons ${l_L}$
are elementary and
the other particles are composite.
The Yukawa coupling of the top quark is
induced by strong gauge dynamics {\eqref{11}}
and this explains the reason why
the Yukawa coupling of top quark $(Y_t)$ is of $O(1)$.
On the other hand,
the Yukawa couplings of bottom quark $(Y_b)$ can not be
derived by the strong gauge dynamics {\eqref{11}}.
${Y_b}$ is put in tree level superpotential
by the hand,
thus
the ${Y_b}$ is suppressed of ${O(\Lambda/M)}$
at the confinement phase.
It can naturally explain
that the ${Y_b}$ becomes smaller than $O(1)$.
In ref.{\cite{luty}},
all quarks and leptons
are composite states of the some elementary particles.
Yukawa couplings of leptons are
induced by non-perturbative effects {\eqref{11}}.
However, it does not follow one philosophy that 
contains a big problem,
that is, the Yukawa couplings of the quarks are
not induced from {\eqref{11}}.

In this article,
we present a composite model in which 
all Yukawa couplings in the SSM  are induced
dynamically.
All quarks, leptons and the Higgs particles
are confining states of preons and anti-preons.
The Yukawa couplings of ordinary quarks and leptons are
induced from the strong gauge dynamics {\eqref{11}}.
Thus this model might suggest the answer of the mystery
of the origin of the Yukawa couplings. 
The anomalies in this model are just the same as the SM
as shown later.
From now on, 
we assume the supersymmetry (SUSY) breaking 
will occurred at low energy. 
And we consider the supersymmetric model 
which is suitable above the SUSY breaking scale.
In this model, there are some unwanted massless fields.
We should consider the mechanism to generate their masses
which does not contradict experiments.

This article is organized as follows.
In section \ref{sec:2}, 
we will focus on a model with one generation,
and we will discuss the properties of this model.
In section \ref{sec:3},
we will try to construct more realistic scenario,
three generation model.
Section \ref{sec:sum} will be devoted to summary and discussion.
\section{One Generation Model}
\label{sec:2}
\setcounter{equation}{0}
{\it Preon Fields}
\par
At first, we focus on a model
with only one generation.
Where quarks and leptons
are confining states of preons and anti-preons.
We introduce new gauge group `` hyper-color'' $SU(7)_H$
in addition to the standard gauge group.
At low energy, the gauge coupling of $SU(7)_H$
becomes strong,
preons and anti-preons confine and
Yukawa couplings
are  dynamically generated
by the non-perturbative effects of $SU(7)_H$.
We consider the gauge group
\beq
SU(7)_H \times SU(3)_C \times SU(2)_L \times U(1)_Y
 \times [U(1)_B \times U(1)_L],
\eqlab{gauge1}
\eeq
where 
$SU(3)_C \times SU(2)_L \times U(1)_Y$
is the ordinary standard model gauge group,
and $U(1)_B$ and $U(1)_L$ are 
baryon and lepton number global symmetries, respectively.
Under the gauge group \eqref{gauge1},
preon and anti-preon fields transform as shown in \tbref{pre}
\begin{center}
\begin{table}[h]
\begin{center}
\begin{tabular}{|c||c|c|c|c||c|c|}
\hline
      & $SU(7)_H$ & $SU(3)_C$ & $SU(2)_L$ & $U(1)_Y$ &$U(1)_B$ & $U(1)_L$\\
\hline
\hline
$\pq$ & $\sb$  & $\sbb$ & $\sb$ & $-{1}/{3}$ & $-{1}/{21}$ & $ {2}/{7}$ \\
$\pl$ & $\sb$  &  $1$   & $\sb$ & $1$        & $ {2}/{7} $ & $-{5}/{7}$ \\
\hline
\hline
$\pu$ & $\sbb$ & $\sb$  & $1$   & $ {4}/{3}$  & ${1}/{21}$ & $-{2}/{7}$ \\
$\pd$ & $\sbb$ & $\sb$  & $1$   & $-{2}/{3}$  & ${1}/{21}$ & $-{2}/{7}$ \\
$\pn$ & $\sbb$ & $1$    & $1$   & $ 0 $       & $-{2}/{7}$ & ${5}/{7}$ \\
$\pe$ & $\sbb$ & $1$    & $1$   & $-2 $       & $-{2}/{7}$ & ${5}/{7}$ \\
\hline
\end{tabular}
\caption {Representation of the preons and anti-preons for the relevant gauge group.}
\tblab{pre}
\end{center}
\end{table}
\end{center}
where $\sb$($\sbb$) denote the fundamental 
(anti-fundamental) representation
of the gauge group.
Assuming that the SM gauge couplings
are weaker than the hyper-color couplings,
this theory is regarded as ${SU(7)_H}$ supersymmetric QCD
with 8 flavors.

Bellow the $SU(7)_H$ confinement scale $\Lambda$,
composite states of ``baryon'' and ``anti-baryon'' appear as \tbref{BB},
\begin{table}[h]
\begin{center}
\begin{tabular}{|c||c|c|c|c||c|c||c|}
\hline
      & $SU(7)_H$ & $SU(3)_C$ & $SU(2)_L$ & $U(1)_Y$ &$U(1)_B$ &
 $U(1)_L$ & baryon $(\times 1/\Lambda^6)$\\
\hline
\hline
$Q$ & $1$ & $\sb$ & $\sb$ & ${1}/{3}$ & ${1}/{3}$ & $ 0 $ & ${\pq^5 \pl^2}$ \\
$l$ & $1$ & $1$   & $\sb$ & $-1$      & $ 0 $     & $ 1 $ & ${\pq^6 \pl}$ \\
\hline
\hline
$\bar{U}$ & $1$ & $\sbb$  & $1$ & $-{4}/{3}$ & $-{1}/{3}$ & $0$&${\pu^2\pd^3\pe\pn}$ \\
$\bar{D}$ & $1$ & $\sbb$  & $1$ & $ {2}/{3}$ & $-{1}/{3}$ & $0$&${\pu^3\pd^2\pe\pn} $ \\
$\bar{N}$ & $1$ & $1$     & $1$ & $ 0 $      & $ 0 $      & $-1$&${\pu^3\pd^3\pe} $ \\
$\bar{E}$ & $1$ & $1$     & $1$ & $ 2 $      & $ 0 $      & $-1$&${\pu^3\pd^3\pn} $ \\
\hline
\end{tabular}
\caption {Transformation of ``baryon'' and ``anti-baryon'' under relevant gauge group}
\tblab{BB}
\end{center}
\end{table}
and states of composite ``meson'' appear as \tbref{M}.
\begin{table}[h]
\begin{center}
\begin{tabular}{|c||c|c|c|c||c|c||c|}
\hline
      & $SU(7)_H$ & $SU(3)_C$ & $SU(2)_L$ & $U(1)_Y$ &$U(1)_B$ &
 $U(1)_L$ & meson $(\times 1/\Lambda)$\\
\hline
\hline
$\Phiu$  &$1$& $1$    &$\sb$&   $ 1$   &    $0$   & $0$&$\pq\pu$ \\
$\Phid$  &$1$& $1$    &$\sb$&   $-1$   &    $0$   & $0$&$\pq\pd$ \\
$\Gu$     &$1$&{\bf Ad.}&$\sb$&   $ 1$   &    $0$   & $0$&$\pq\pu$ \\
$\Gd$     &$1$&{\bf Ad.}&$\sb$&   $-1$   &    $0$   & $0$&$\pq\pd$ \\
$X$      &$1$& $\sbb$ &$\sb$&$-{1}/{3}$&$-{1}/{3}$& $1$&$\pq\pn$ \\
$\bar{X}$&$1$& $\sb$  &$\sb$&$ {1}/{3}$&$ {1}/{3}$&$-1$&$\pl\pd$ \\
$Y$      &$1$& $\sbb$ &$\sb$&$-{7}/{3}$&$-{1}/{3}$& $1$&$\pq\pe$ \\
$\bar{Y}$&$1$& $\sb$  &$\sb$&$ {7}/{3}$&$ {1}/{3}$&$-1$&$\pl\pu$ \\
$\phiu$  &$1$& $1$    &$\sb$&   $ 1$   &    $0$   & $0$&$\pl\pn$ \\
$\phid$  &$1$& $1$    &$\sb$&   $-1$   &    $0$   & $0$&$\pl\pe$ \\
\hline
\end{tabular}
\caption {Transformation of ``meson'' under relevant gauge group}
\tblab{M}
\end{center}
\end{table}

In addition to the ordinary one generation 
quarks and leptons in the SSM
($Q$, $\bar{U}$, $\bar{D}$, $\ell$ and $\bar{E}$),
there exist
right-handed neutrino ($\bar{N}$) and
two set of the Higgs doublet
($\bar{\Phi}_{\rm u}$, $\bar{\Phi}_{\rm d}$ and 
 $\bar{\phi}_{\rm u}$, $\bar{\phi}_{\rm d}$),
color octet fields ($G_{\rm u}$ and $G_{\rm d}$)
and two sets of the leptoquarks
($X$, $\bar{X}$ and $Y$, $\bar{Y}$)
as the $SU(7)_H$ ``meson''.
This model also predicts the existences of
right-handed neutrino ($\bar{N}$),
two set of the Higgs doublet
($\bar{\Phi}_{\rm u}$, $\bar{\Phi}_{\rm d}$ and 
 $\bar{\phi}_{\rm u}$, $\bar{\phi}_{\rm d}$).

Bellow the confinement scale $\Lambda$,
quarks, leptons and Higgs particles
appear as ``baryon (anti-baryon)'' and ``meson'', respectively.
The dynamical superpotential {\eqref{11}} becomes
\bea
W_{\rm dyn} &=& y_u\bar{U}(\Phiu + \phiu)Q
               + y_d\bar{D}(\Phid + \phid)Q \nn \\
            & &+ y_n\bar{N}(\Phiu + \phiu) \ell
               + y_e\bar{E}(\Phid + \phid) \ell \nn \\
            & &+ {\alpha_u} \bar{U} G_{\rm u} Q 
               + {\alpha_d} \bar{D} G_{\rm d} Q \nn \\
            & &+ {\beta_u} \bar{U} \bar{Y} \ell
               + {\beta_d} \bar{D} \bar{X} \ell
               + {\beta_n} \bar{N}   X     Q
               + {\beta_e} \bar{E}   Y     Q \nn \\
            & &- \mbox{determinant},
\eqlab{Wdy}
\eea
where determinant denotes 8 powers interactions
of composite meson fields.
And all of coefficient,
$y_u, y_d, y_n, y_e,
 \alpha_u, \alpha_d,
 \beta_u, \beta_d, \beta_n$ and $\beta_d$,
are of $O(1)$ and equal
if $SU(3)_C \times SU(2)_L \times U(1)_Y$ is absent.
Above the $SU(7)_H$ confinement scale $\Lambda$,
this model do not have any Yukawa couplings,
however when the $SU(7)_H$ gauge coupling becomes
strong,
all Yukawa couplings in the SSM emerge from
the confining dynamics.
\\
\\
{\it Anomalies and Property}
\par
In this model anomalies are absent
except those of
\bea
SU^2(2)_{L} \times U(1)_{B+L},
\mbox{\qquad \qquad}
 U^2(1)_{Y} \times U(1)_{B+L}.
\eqlab{globalan}
\eea
These anomalies are the exactly same as those of the 
SSM{\footnote {The former one is consider to be useful for electro-weak baryogenesis \cite{bg}.}}.

If we combine ($\pu$ and $\pd$) and ($\pn$ and $\pe$)
into $SU(2)_R$ doublet,
this model can be embedded into a Left-Right symmetric model.
And, if preon fields (${\pq}$ and ${\pl}$) and 
anti-preon fields (${\pu}$, ${\pd}$, ${\pe}$ and ${\pn}$)
are combined into ${SU(4) \times SU(2)_L \times SU(2)_R}$ fields,
this model can be extended to
the model with Pati-Salam type gauge symmetry.
\cite{PS}.
\\
\\
{\it Higher order interaction}
\par
The Yukawa interaction \eqref{Wdy} does not give masses to 
color octet fields $G_{\rm u}$ and $G_{\rm d}$,
leptoquarks $X$, $\bar{X}$, $Y$ and $\bar{Y}$,
and Higgs particles $\Phid$, $\Phiu$, $\phid$ and $\phiu$.
Thus, we must consider the mechanism to generate their masses.
For this, we introduce 4 dimensional higher order
operators in the tree level superpotential.
The general gauge invariant dimension 4 operators are
\beq
\tilde{W}_{tree} = g_{G}   \dfrac{\pq\pq\pu\pd}{M}
                 + g_{\mu} \dfrac{\pl\pl\pe\pn}{M}
                 + g_{x}   \dfrac{\pq\pl\pd\pn}{M}
                 + g_{y}   \dfrac{\pq\pl\pu\pe}{M},
\eqlab{prepot}
\eeq
where $M$ is a mass parameter
and we assume that coefficient
${g_{\mu}}$, ${g_G}$, ${g_x}$
and ${g_y}$ are of ${O(1)}$.
Bellow the energy scale $\Lambda$,
\eqref{prepot} becomes
\beq
W_{tree}^{eff} =m(g_G G_{\rm u}G_{\rm d} + g_G \Phiu \Phid 
               + g_\mu \phiu \phid + g_x \bar{X} X
               + g_x \phiu \Phid + g_y \bar{Y} Y + g_y \Phiu \phid),
\eqlab{Weff}
\eeq
where mass parameter $m$ define as
\beq
m \equiv \dfrac{\Lambda^2}{M}.
\eqlab{def_m}
\eeq
Therefore, leptoquarks and color-octet particles have mass
of $O(m)$.
The mass matrix of Higgs particles becomes
\beq
\left(
\begin{array}{cc}
\bar{\phi}_{\rm u} \mbox{ } \bar{\Phi}_{\rm u}
\end{array}
\right)
\left( \begin{array}{cc}
  m g_\mu &  m g_x \\
  m g_y   &  m g_G 
\end{array} \right)
\left( \begin{array}{c}
\bar{\phi}_{\rm d} \\
\bar{\Phi}_{\rm d}
\end{array} \right).
\eqlab{matrix_higgs}
\eeq
If $g_{\mu}g_G \simeq g_x g_y$,
one of the Higgs masses become small.

\section{Three Generation Model}
\label{sec:3}
\setcounter{equation}{0}
\hspace*{\parindent}
Now, let us try to extend a previous model
to the three generation model.
We may consider following three possibilities:
\\
\\
{\bf{\it i) Family Symmetry}}
\par
We add the global family symmetry
${SU(3)_F}$ to \eqref{gauge1}.
We adapt the gauge symmetry,
$SU(23)_H \times SU(3)_C \times SU(2)_L \times U(1)_Y
\times [SU(3)_F \times U(1)_B \times U(1)_L]$.

Under the gauge symmetry preons and anti-preons transform as \tbref{3pre}.
Baryon (anti-baryon) states are fundamental (anti-fundamental)
representation of the family symmetry $SU(3)_F$ (see \tbref{3BB}).
It is worth noting that $SU(3)_F$ octet 
``meson'' fields are added
to the one generation model (see \tbref{3M}).
Bellow the composite scale $\Lambda$,
dynamically induced superpotential {\eqref{11}}
is generated:
\beq
W_{\rm dyn} = W_{\rm dyn}^{singlet} + W_{\rm dyn}^{octet},
\eeq
where
\bea
W_{\rm dyn}^{singlet} 
      &=& y_{u1}\bar{U}(\Phiu^{\rm s} + \phiu^{\rm s})Q
         + y_{d1}\bar{D}(\Phid^{\rm s} + \phid^{\rm s})Q \nn \\
      & &+ y_{n1}\bar{N}(\Phiu^{\rm s} + \phiu^{\rm s}) \ell
         + y_{e1}\bar{E}(\Phid^{\rm s} + \phid^{\rm s}) \ell \nn \\
      & &+ {\alpha_{u1}} \bar{U} G_{\rm u}^{\rm s} Q 
         + {\alpha_{d1}} \bar{D} G_{\rm d}^{\rm s} Q \nn \\
      & &+ {\beta_{u1}} \bar{U} \bar{Y}^{\rm s} \ell
         + {\beta_{d1}} \bar{D} \bar{X}^{\rm s} \ell
         + {\beta_{n1}} \bar{N}   X^{\rm s}     Q
         + {\beta_{e1}} \bar{E}   Y^{\rm s}     Q,
\eqlab{Wdy-F1}
\eea
and
\bea
W_{\rm dyn}^{octet}
      &=& y_{u8}\bar{U}(\Phiu^{\rm o} + \phiu^{\rm o})Q
         + y_{d8}\bar{D}(\Phid^{\rm o} + \phid^{\rm o})Q \nn \\
      & &+ y_{n8}\bar{N}(\Phiu^{\rm o} + \phiu^{\rm o}) \ell
         + y_{e8}\bar{E}(\Phid^{\rm o} + \phid^{\rm o}) \ell \nn \\
      & &+ {\alpha_{u8}} \bar{U} G_{\rm u}^{\rm o} Q 
         + {\alpha_{d8}} \bar{D} G_{\rm d}^{\rm o} Q \nn \\
      & &+ {\beta_{u8}} \bar{U} \bar{Y}^{\rm o} \ell
         + {\beta_{d8}} \bar{D} \bar{X}^{\rm o} \ell
         + {\beta_{n8}} \bar{N}   X^{\rm o}     Q
         + {\beta_{e8}} \bar{E}   Y^{\rm o}     Q.
\eqlab{Wdy-F8}
\eea
Here we neglect the determinant term for mesons.
All coefficient are considered be of ${O(1)}$, and
suffix of the particles
``s'' and ``o'' denote the singlet and the octet 
representation of the $SU(3)_F$ family symmetry, respectively.

This model has the family symmetry
which is troublesome.
For example,
flavor changing neutral current (FCNC) exists.
Also Nambu-Goldstone (NG) bosons appears 
when family symmetry is broken.
When the global family symmetry is
spontaneously broken,
many massless NG bosons appear.
We need to consider how to break the flavor symmetry
to make realistic scenario.
\\
\\
{\bf{\it ii) Three Copies of One Generation}}
\par
The second way is to consider three copies of 
one generation model.
The gauge group is extended to,
$ SU(7)_{H_1} \times SU(7)_{H_2} \times SU(7)_{H_3} \times
  SU(3)_C \times SU(2)_L \times U(1)_Y \times
  [U(1)_B \times U(1)_L \times {\mbox {\bf Z}}_8]$,
where three $SU(7)_{H_i}$ are ``hyper-color'' gauge groups,
by which dynamics preons and anti-preons
of each generation are confined.
Bellow the confinement scale of the all hyper-color dynamics,
the composite fields appear as shown in
\tbref{3BB2} and \tbref{3M2}, 
and
the superpotential is derived by the non-perturbative effects as
\bea
W_{\rm dyn} = &\displaystyle{\sum_{a=1}^3}&
   \left\{ 
          y_{ua}\bar{U_a}(\Phiu + \phiu)_a Q_a
        + y_{da}\bar{D_a}(\Phid + \phid)_a Q_a 
   \right .
   \nn \\
         & & 
   \left .
        + y_{na}\bar{N_a}(\Phiu + \phiu)_a \ell_a
        + y_{ea}\bar{E_a}(\Phid + \phid)_a \ell_a
   \right\},
\eqlab{Wdy-3}
\eea
This model predicts that
each generation has two sets of the Higgs doublets,
and the Yukawa couplings of the first, the second and the third
generations are of the ${O(1)}$.
Since ${W_{\rm dyn}}$,
\eqref{Wdy-3} does not include flavor mixing interactions,
we introduce the higher dimensional operators in the tree level
superpotential.
\bea
W_{\rm tree} =&\displaystyle{\sum_{i,j,k=1}^3}&
   \left\{ 
          y^u_{ijk} (\dfrac{\Lambda}{M})^{13}
            \bar{U_i}(\Phiu + \phiu)_k Q_j
         +y^d_{ijk} (\dfrac{\Lambda}{M})^{13}
            \bar{D_i}(\Phid + \phid)_k Q_j
   \right .
   \nn \\
         & & 
   \left .
          y^n_{ijk} (\dfrac{\Lambda}{M})^{13}
            \bar{N_i}(\Phiu + \phiu)_k \ell_j
         +y^e_{ijk} (\dfrac{\Lambda}{M})^{13}
            \bar{E_i}(\Phid + \phid)_k \ell_j
   \right\}
\eqlab{Wdy-H3}
\eea
in order to obtain the KM matrix.
Where $y^{a}_{ijk}={\alpha}^{a}_{ijk}
{\Lambda_i^6 \Lambda_k \Lambda_j^6}/{\Lambda^{13}}$
($\alpha_{ijk}^{a} \simeq O(1)$) and 
$\Lambda_i$ is the confinement scale of $SU(7)_{H_{i}}$.
The ${{\mbox {\bf Z}}_8}$ symmetry is introduce in order to
avoid large FCNC.
Here preons and anti-preons are assigned to have charge 
\bea
P_1 : 1,      \mbox{\quad} P_2 : 3,      \mbox{\quad}  P_3 : 5, \nn \\
\bar{P}_1 : 6,\mbox{\quad} \bar{P}_2 : 5,\mbox{\quad}  \bar{P}_3 : 4,
\eqlab{z8}
\eea
under the ${{\mbox {\bf Z}}_8}$ symmetry.
Here $P_i$ ($\bar{P}_i$) is denote 
{\it i} th generation preons (anti-preons) (see \tbref{3pre2}).
Texture of the Yukawa couplings are
\beq
Y^{\rm{u,d,e,n}} \sim
\bmaT
   O(1)  &   0     &  0        \\
\ve^{13} &  O(1)   & \ve^{13}  \\
    0    &  0      & O(1)
\emaT,
\eqlab{mass-3}
\eeq
in the flavor space,
where $\ve \sim \dfrac{\Lambda}{M}$.

The $(1,2)$, $(1,3)$, $(3,1)$ and $(3,2)$ elements of all Yukawa coupling
matrices vanish because of {\bf Z}$_8$ symmetry.
Eq.(\ref{eqn:mass-3}) is the texture of the Yukawa couplings
and not the mass matrices.
It is necessary to determine the SUSY breaking terms
to deduce the VEV's of Higgs particles in each generation.
\\
\\
{\bf {\it iii) 2-1 Generation}}
\par
The last one is that only third generation and Higgs particles
to be composite,
and quarks and leptons of the first and the second generations
are elementally particles.
This model include two set of Higgs doublets,
which is different from models {\it i}) and {\it ii}).
As for the first and the second generations,
the quarks and leptons are singlet of the ${SU(7)_H}$.
On the other hand,
the preons and anti-preons of the third generation
that have an opposite charge of
the first and the second generations for the SM charge
confine to be composite states.
Fields with an opposite charge,
so-called ``anti-generation(miller field)'',
is naturally included in the String Inspired Model
{\cite{SIM}}.

When $SU(7)_H$ gauge coupling becomes strong,
dynamical superpotential \eqref{Wdy} are induced.
Only the Yukawa couplings of third generations
are induced.
Thus, at the tree level
we should introduce the Yukawa couplings of 
the first and second generations,
and
the \eqref{prepot} gives masses
of Higgs particles, leptoquarks and color-octet Higgs.
We introduce the tree level superpotential as
\bea
{W}_{tree}
 &=& y_u^{ij} \bar{u}_i \dfrac{(\pq\pu + \pl\pn)}{M} q_j
    + y_d^{ij} \bar{d}_i \dfrac{(\pq\pd + \pl\pe)}{M} q_j \nn \\
 & &+ y_u^{i3} \bar{u}_i 
         \dfrac{(\pq\pu + \pl\pn)\pq^5 \pl^2}{M^7}
    + y_d^{i3}\bar{d}_i 
         \dfrac{(\pq\pd + \pl\pe)\pq^5 \pl^2}{M^7} \nn \\
 & &+ y_u^{3j}
         \dfrac{\pu^2\pd^3\pe\pn(\pq\pu + \pl\pn)}{M^7} q_j
    +  y_d^{3j}
         \dfrac{\pd^3\pu^2\pe\pn(\pq\pd + \pl\pe)}{M^7} q_j \nn \\
 & &+ y_n^{ij} \bar{n}_i \dfrac{(\pq\pu + \pl\pn)}{M} l_j
    + y_e^{ij} \bar{e}_i \dfrac{(\pq\pd + \pl\pe)}{M} l_j \nn \\
 & &+ y_n^{i3}\bar{n}_i 
         \dfrac{(\pq\pu + \pl\pn)\pq^6 \pl}{M^7}
    +  y_e^{i3}\bar{e}_i 
         \dfrac{(\pq\pd + \pl\pe)\pq^6 \pl}{M^7} \nn \\
 & &+ y_n^{3j}
        \dfrac{\pu^3\pd^3\pe(\pq\pu + \pl\pn)}{M^7} l_j
    +  y_e^{3j}
        \dfrac{\pd^3\pu^3\pn(\pq\pd + \pl\pe)}{M^7} l_j \nn \\
 & &+ g_{G}   \dfrac{\pq\pq\pu\pd}{M}
    + g_{\mu} \dfrac{\pl\pl\pe\pn}{M}
    + g_{x}   \dfrac{\pq\pl\pd\pn}{M}
    + g_{y}   \dfrac{\pq\pl\pu\pe}{M},
\eqlab{W1pre}
\eea
where
$\bar{u}_i$, $\bar{d}_i$, $q_i$, $\bar{e}_i$, $\bar{n}_i$ and $\ell_i$
are $SU(7)_H$ singlet 
up-part, down-part, doublet quark,
charged lepton, neutral lepton and doublet lepton 
of {\it i} th generation $(i,j = 1,2)$, respectively.
${M}$ is the arbitrary mass scale of this model.

Bellow the confinement scale $\Lambda$,
the effective superpotential
${W^{(eff)}}$ becomes
\bea
W^{(eff)}
 &=& W^{(eff)}_{tree} + W_{\rm dyn} \nn \\
 &=&  y_u^{ij} \ve   \bar{u}_i (\Phiu+\phiu) q_j
    + y_u^{i3} \ve^7 \bar{u}_i (\Phiu+\phiu) Q
    + y_u^{3j} \ve^7 \bar{U}   (\Phiu+\phiu) q_j  \nn \\
 & &+ y_d^{ij} \ve   \bar{d}_i (\Phid+\phid) q_j
    + y_d^{i3} \ve^7 \bar{d}_i (\Phid+\phid) Q
    + y_d^{3j} \ve^7 \bar{D}   (\Phid+\phid) q_j  \nn \\
 & &+ y_n^{ij} \ve   \bar{n}_i (\Phiu+\phiu) l_j
    + y_n^{i3} \ve^7 \bar{n}_i (\Phiu+\phiu) \ell
    + y_n^{3j} \ve^7 \bar{N}   (\Phiu+\phiu) l_j  \nn \\
 & &+ y_e^{ij} \ve   \bar{e}_i (\Phid+\phid) l_j
    + y_e^{i3} \ve^7 \bar{e}_i (\Phid+\phid) \ell
    + y_e^{3j} \ve^7 \bar{E}   (\Phid+\phid) l_j  \nn \\
 & &+{\alpha_u}^{ij} \ve   \bar{u}_i G_{\rm u} q_j
    +{\alpha_u}^{i3} \ve^7 \bar{u}_i G_{\rm u} Q
    +{\alpha_u}^{3j} \ve^7 \bar{U}   G_{\rm u} q_j  \nn \\
 & &+{\alpha_d}^{ij} \ve   \bar{d}_i G_{\rm d} q_j
    +{\alpha_d}^{i3} \ve^7 \bar{d}_i G_{\rm d} Q
    +{\alpha_d}^{3j} \ve^7 \bar{D}   G_{\rm d} q_j  \nn \\
 & &+  y_u\bar{U}(\Phiu + \phiu)Q
    + y_d\bar{D}(\Phid + \phid)Q \nn \\
 & &+ y_n\bar{N}(\Phiu + \phiu) \ell
    + y_e\bar{E}(\Phid + \phid) \ell \nn \\
 & &+ {\alpha_u} \bar{U} G_{\rm u} Q 
    + {\alpha_d} \bar{D} G_{\rm d} Q \nn \\
 & &+ {\beta_u} \bar{U} \bar{Y} \ell
    + {\beta_d} \bar{D} \bar{X} \ell
    + {\beta_n} \bar{N}   X     Q
    + {\beta_e} \bar{E}   Y     Q \nn \\
 & &+ m(g_G G_{\rm u}G_{\rm d} + g_G \Phiu \Phid 
    + g_\mu \phiu \phid + g_x \bar{X} X
    + g_x \phiu \Phid + g_y \bar{Y} Y + g_y \Phiu \phid) \nn \\
 & &- \mbox{determinant of mesons}, 
\eqlab{W1eff}
\eea
where $\ve \equiv \dfrac{\Lambda}{M}$ and
$m \equiv \dfrac{\Lambda^2}{M}$.
When ${\Phiu+\phiu}$ and ${\Phid+\phid}$ 
have vacuum expectation values (VEVs),
mass matrices in the flavor space become
\beq
M_{\rm{u,d,e,n}} \sim
\bmaT
\ve   & \ve   & \ve^7 \\
\ve   & \ve   & \ve^7 \\
\ve^7 & \ve^7 &   1
\emaT,
\eqlab{mass-2-1}
\eeq
where each element of mass matrices has coefficient
$y^{ij}_{a}$, $(i,j = 1\sim3$ and $a=u,d,n,e)$.
\\
\\
\par
Now let us consider the phenomenology.
We set the confinement scale at $\Lambda \simeq O(10^{17})$ GeV
because the $O(1)$ Yukawa coupling of
third generations can be realized in $SO(10)$ GUT model \cite{so10GUT}.
We consider $M$ as the Plank scale,
then, it becomes ${\ve \simeq O(10^{-1})}$,
and $m$ can be estimated:
\beq
m \sim \dfrac{\Lambda^2}{M} \sim 10^{16} \mbox{GeV}.
\eeq
Therefore leptoquark ${X, \bar{X}, Y}$ and ${\bar{Y}}$
and $SU(3)_C$ octet fields ${\Gd}$, ${\Gu}$ have large mass
of ${O(m)}$.
However,
masses of each element of the Higgs mass matrix,
\eqref{matrix_higgs},
are also of ${O(m)}$.
In order to obtain the realistic ${\mu}$-term
of ${O(M_z)}$,
we have to fine tune couplings of \eqref{Weff}.
For example,
when we fine tune the ${g_{\mu}g_{G} = g_{x}g_{y}}$
in \eqref{matrix_higgs},
the determinant of the \eqref{matrix_higgs}
becomes to be zero.
In this case, by introducing another gauge group,
we could obtain the suitable ${\mu}$-term \cite{yanagida}.
\\
\\
\par
When Yukawa couplings of the third generation
are united at the confinement scale ${\Lambda \sim 10^{17}}$ GeV
{\footnote{${y_b}$, ${y_{\tau}}$, ${y_t}$ and ${y_n}$ are
Yukawa coupling of the b quark, ${\tau}$ lepton, 
t quark and ${\nu_{\tau}}$, respectively}}.
\beq
y_b \simeq y_{\tau} \simeq y_t \simeq y_n,
\eeq
the relationships between the Yukawa couplings of
the third generation become
\beq
y_b \simeq 3 y_{\tau} \simeq y_t \simeq 3y_n
\eqlab{yyy}
\eeq
at the weak scale ${O(M_z)}$,
where we tune the value of ${\tan \beta}$
to obtain the difference between
the top quark and bottom quark mass.
In this case, however, ${\nu_{\tau}}$ becomes heavy
because neutrino and top quark couple the same Higgs particles
and there is the relation {\eqref{yyy}}.
Mass of the neutrino (${m_{\nu}}$) become
about 1/3 of the top mass,
that is 
\beq
m_{\nu} \simeq 60 {\mbox { GeV.}}
\eqlab{mass_neu}
\eeq
It is not in agreement with the nature.
Thus,
we must introduce a Majorana type mass matrix
for the right handed neutrinos.
Mass of neutrino becomes small
by the See-Saw mechanism {\cite{SeeSaw}}
and the mass of ${\nu_{\tau}}$ does not contradict
then experiments.
We introduce new elementary gauge singlet fields $S$
with $-2$ lepton number.
We also add discrete ${\mbox{\bf Z}}_3$ symmetry,
and we assign for the neutrinos and singlet ${S}$ as
\beq
\mbox{\bf Z}_3(n_1)=1, {\mbox{\quad}}
\mbox{\bf Z}_3(n_2)=2, {\mbox{\quad}}
\mbox{\bf Z}_3(S)  =1,
\eqlab{Singlet}
\eeq
where $n_i$ is {\it i} th generation neutrino.
The other elementary particles do not have 
this discrete ${\mbox{\bf Z}}_3$ charge.
Thus, the new additional superpotential is
\beq
W_{\nu}= S n_1 n_1 + S n_2 N \left(\dfrac{\Lambda}{M} \right)^6.
\eqlab{W_nu}
\eeq
Then the mass matrix of right-hand neutrino becomes
\beq
M_N =
\bmaT
   S    &   0     & 0 \\
   0    &   0     & S \ve^6 \\
   0    & S \ve^6 & 0
\emaT.
\eeq
When the singlet particle $S$ has vacuum expectation value
$<S> \sim 10^{16}$ GeV, all neutrino mass become light
enough to be consistent with experiments.
\section{Summary}
\label{sec:sum}
\setcounter{equation}{0}
\hspace*{\parindent}
We present a composite model 
that is based on non-perturbative effects
of ${N=1}$ supersymmetric $SU(N_C)$ gauge
theory with $N_f=N_C+1$ flavors.
In this model, we consider ${N_C = 7}$,
where all particles in the SSM are composite states of
elementary preons and anti-preons.
When ${SU(7)_H}$ hyper-color gauge couplings become strong,
preons and anti-preons are confined to be
quarks, leptons and Higgs particles.
At the same time,
Yukawa interactions in the SSM emerge from
the non-perturbative dynamics of the ${SU(7)_H}$ hyper-color. 

At first, we consider one generation model.
This model predicts the existence of 
two sets of Higgs doublets.
However, unwanted massless fields also appear,
which are leptoquarks and color-octet particles.
We introduce 4 dimensional higher order operators
in the superpotential to generate the
masses of these unwanted particles.

We then generalize a model to three generations
in three ways.
In the model {\it i}),
we introduce global family symmetry ${SU(3)_F}$.
In this model, many Higgs particles are induced
by the strong gauge dynamics.
Especially,
this model include octet 
Higgs particles of the ${SU(3)_F}$ family symmetry.
Therefor, it is necessary to think about
more complex and newer scenario
to make this model more realistic.
In Model {\it ii}), we introduce 
three sets of hyper-color gauge symmetry
which confine each generations.
This model predicts that each generation has
two sets of the Higgs doublets,
and the Yukawa couplings of the all generations 
are of ${O(1)}$.
Since flavor mixing matrix is not induced
by the gauge dynamics,
we introduce tree level the higher dimensional 
operators.
At this time we introduce the {\bf Z${_8}$}
symmetry in order to avoid large FCNC.
In model {\it iii}),
the first and the second generations are
hyper-color singlet elementary particles
and only third generations and Higgs Particles
are composite states of the preons and anti-preons.
We should adjust confinement scale to ${10^{17}}$ GeV
because the all Yukawa couplings of the third generation
are of ${O(1)}$.
Then, we have to fine tune the couplings
in the \eqref{mass-3}
to obtain a realistic ${\mu}$-term.
And we introduce the new elementary gauge singlet fields $S$
in order to obtain small masses of neutrinos.

In this article we have suggested one of the possibilities 
for the origin of Yukawa coupling.
Further work is necessary to understand to determine
the mechanism of SUSY breaking and the electroweak symmetry breaking
in order to discuss the mass matrices and flavor mixing.
We hope that these models should shed some light on
the origin of Yukawa couplings.
\section*{Acknowledgments}
We would like to thank Prof. A. I. Sanda 
for fruitful comments
and careful reading of manuscripts.
One of the author (N. H. ) is grateful to 
Prof. Y. Okada for helpful discussions and comments.

\newpage
\appendix
\renewcommand{\thefootnote}{\arabic{footnote}}
\setcounter{footnote}{0}
\setcounter{table}{0}
\section{Gauge representations of the each three generation model}
\bfoot
\begin{table}[h]
\begin{center}
\begin{tabular}{|c||c|c|c|c||c|c|c|}
\hline
      & $SU(23)_H$ & $SU(3)_C$ & $SU(2)_L$ & $U(1)_Y$ & $U(1)_F$ & $U(1)_B$ & $U(1)_L$\\
\hline
\hline
$\pq$&$\sb$ &$\sbb$&$\sb$&$-{1}/{3}$&$\sb$ &$-{5}/{69}$&$ {6}/{23}$ \\
$\pl$&$\sb$ & $1$  &$\sb$&$1$       &$\sb$ &$ {6}/{23}$&$-{7}/{23}$ \\
\hline
\hline
$\pu$&$\sbb$&$\sb$ & $1$ &$ {4}/{3}$&$\sbb$&$ {5}/{69}$&$-{6}/{23}$ \\
$\pd$&$\sbb$&$\sb$ & $1$ &$-{2}/{3}$&$\sbb$&$ {5}/{69}$&$-{6}/{23}$ \\
$\pn$&$\sbb$& $1$  & $1$ &$ 0 $     &$\sbb$&$-{6}/{23}$&$ {7}/{23}$ \\
$\pe$&$\sbb$& $1$  & $1$ &$-2 $     &$\sbb$&$-{6}/{23}$&$ {7}/{23}$ \\
\hline
\end{tabular}
\caption {Preon and anti-preon fields in Model {\it i})}
\tblab{3pre}
\end{center}
\end{table}
\begin{table}[h]
\begin{center}
\begin{tabular}{|c||c|c|c|c||c|c|c||c|}
\hline
      & $SU(23)_H$ & $SU(3)_C$ & $SU(2)_L$ & $U(1)_Y$ & $SU(3)_F$ &$U(1)_B$ &
 $U(1)_L$ & baryon states $(\times 1/\Lambda^{22})$\\
\hline
\hline
$Q$ &1& $\sb$ & $\sb$ & ${1}/{3}$ &1& ${1}/{3}$ & $ 0 $ & ${\pq^{17} \pl^6}$ \\
$l$ &1& $1$   & $\sb$ & $-1$      &1& $ 0 $     & $ 1 $ & ${\pq^{18} \pl^5}$ \\
\hline
\hline
$\bar{U}$&1& $\sbb$  & $1$ & $-{4}/{3}$ &1& $-{1}/{3}$ & $0$&${\pu^8\pd^9\pe^3\pn^3}$ \\
$\bar{D}$&1& $\sbb$  & $1$ & $ {2}/{3}$ &1& $-{1}/{3}$ & $0$&${\pu^9\pd^8\pe^3\pn^3} $ \\
$\bar{N}$&1& $1$     & $1$ & $ 0 $      &1& $ 0 $      & $-1$&${\pu^9\pd^9\pe^3\pn^2} $ \\
$\bar{E}$&1& $1$     & $1$ & $ 2 $      &1& $ 0 $      & $-1$&${\pu^9\pd^9\pe^2\pn^3} $ \\
\hline
\end{tabular}
\caption {``Baryon'' and ``anti-baryon'' fields in Model {\it i})}
\tblab{3BB}
\end{center}
\end{table}
\begin{table}[h]
\begin{center}
\begin{tabular}{|c||c|c|c|c||c|c|c||c|}
\hline
      & $SU(7)_H$ & $SU(3)_C$ & $SU(2)_L$ & $U(1)_Y$ & $SU(3)_F$ &$U(1)_B$ &
 $U(1)_L$ & meson states $(\times 1/\Lambda)$\\
\hline
\hline
$\Phiu^{\rm s}$  &$1$& $1$    &$\sb$&   $ 1$   &1&    $0$   & $0$&$\pq\pu$ \\
$\Phid^{\rm s}$  &$1$& $1$    &$\sb$&   $-1$   &1&    $0$   & $0$&$\pq\pd$ \\
$\Gu^{\rm s}$    &$1$&{\bf Ad}&$\sb$&   $ 1$   &1&    $0$   & $0$&$\pq\pu$ \\
$\Gd^{\rm s}$    &$1$&{\bf Ad}&$\sb$&   $-1$   &1&    $0$   & $0$&$\pq\pd$ \\
$X^{\rm s}$      &$1$& $\sbb$ &$\sb$&$-{1}/{3}$&1&$-{1}/{3}$& $1$&$\pq\pn$ \\
$\bar{X}^{\rm s}$&$1$& $\sb$  &$\sb$&$ {1}/{3}$&1&$ {1}/{3}$&$-1$&$\pl\pd$ \\
$Y^{\rm s}$      &$1$& $\sbb$ &$\sb$&$-{7}/{3}$&1&$-{1}/{3}$& $1$&$\pq\pe$ \\
$\bar{Y}^{\rm s}$&$1$& $\sb$  &$\sb$&$ {7}/{3}$&1&$ {1}/{3}$&$-1$&$\pl\pu$ \\
$\phiu^{\rm s}$  &$1$& $1$    &$\sb$&   $ 1$   &1&    $0$   & $0$&$\pl\pn$ \\
$\phid^{\rm s}$  &$1$& $1$    &$\sb$&   $-1$   &1&    $0$   & $0$&$\pl\pe$ \\
\hline
\hline
$\Phiu^{\rm o}$  &$1$& $1$    &$\sb$&   $ 1$   &{\bf Ad}&    $0$   & $0$&$\pq\pu$ \\
$\Phid^{\rm o}$  &$1$& $1$    &$\sb$&   $-1$   &{\bf Ad}&    $0$   & $0$&$\pq\pd$ \\
$\Gu^{\rm o}$    &$1$&{\bf Ad}&$\sb$&   $ 1$   &{\bf Ad}&    $0$   & $0$&$\pq\pu$ \\
$\Gd^{\rm o}$    &$1$&{\bf Ad}&$\sb$&   $-1$   &{\bf Ad}&    $0$   & $0$&$\pq\pd$ \\
$X^{\rm o}$      &$1$& $\sbb$ &$\sb$&$-{1}/{3}$&{\bf Ad}&$-{1}/{3}$& $1$&$\pq\pn$ \\
$\bar{X}^{\rm o}$&$1$& $\sb$  &$\sb$&$ {1}/{3}$&{\bf Ad}&$ {1}/{3}$&$-1$&$\pl\pd$ \\
$Y^{\rm o}$      &$1$& $\sbb$ &$\sb$&$-{7}/{3}$&{\bf Ad}&$-{1}/{3}$& $1$&$\pq\pe$ \\
$\bar{Y}^{\rm o}$&$1$& $\sb$  &$\sb$&$ {7}/{3}$&{\bf Ad}&$ {1}/{3}$&$-1$&$\pl\pu$ \\
$\phiu^{\rm o}$  &$1$& $1$    &$\sb$&   $ 1$   &{\bf Ad}&    $0$   & $0$&$\pl\pn$ \\
$\phid^{\rm o}$  &$1$& $1$    &$\sb$&   $-1$   &{\bf Ad}&    $0$   & $0$&$\pl\pe$ \\
\hline
\end{tabular}
\caption { ``Meson'' fields in Model {\it i})}
\tblab{3M}
\end{center}
\end{table}
\begin{table}[h]
\begin{center}
\begin{tabular}{|c||c|c|c|c|c|c||c|c|c|}
\hline
      & $SU(7)_{H_1}$ & $SU(7)_{H_2}$ & $SU(7)_{H_3}$ & $SU(3)_C$ &
 $SU(2)_L$ & $U(1)_Y$ &$U(1)_B$ & $U(1)_L$ & {\bf Z}$_8$\\
\hline
\hline
$\pq$$_1$ & $\sb$&1&1& $\sbb$&$\sb$&$-{1}/{3}$ & $-{1}/{21}$&$ {2}/{7}$&$1$\\
$\pl$$_1$ & $\sb$&1&1&  $1$  &$\sb$&$1$        & $ {2}/{7} $&$-{5}/{7}$&$1$\\
\hline
$\pu$$_1$&$\sbb$&1&1&$\sb$&$1$  &$ {4}/{3}$&${1}/{21}$ &$-{2}/{7}$&$6$ \\
$\pd$$_1$&$\sbb$&1&1&$\sb$&$1$  &$-{2}/{3}$&${1}/{21}$ &$-{2}/{7}$&$6$ \\
$\pn$$_1$&$\sbb$&1&1&$1$  &$1$  &$ 0 $     &$-{2}/{7}$ &${5}/{7}$ &$6$\\
$\pe$$_1$&$\sbb$&1&1&$1$  &$1$  &$-2 $     &$-{2}/{7}$ &${5}/{7}$ &$6$\\
\hline
\hline
$\pq$$_2$&1&$\sb$&1&$\sbb$&$\sb$&$-{1}/{3}$&$-{1}/{21}$&$ {2}/{7}$&$3$\\
$\pl$$_2$&1&$\sb$&1& $1$  &$\sb$&$1$       &$ {2}/{7} $&$-{5}/{7}$&$3$\\
\hline
$\pu$$_2$&1&$\sbb$&1&$\sb$&$1$  &$ {4}/{3}$&${1}/{21}$ &$-{2}/{7}$&$5$ \\
$\pd$$_2$&1&$\sbb$&1&$\sb$&$1$  &$-{2}/{3}$&${1}/{21}$ &$-{2}/{7}$&$5$ \\
$\pn$$_2$&1&$\sbb$&1&$1$  &$1$  &$ 0 $     &$-{2}/{7}$ &${5}/{7}$ &$5$\\
$\pe$$_2$&1&$\sbb$&1&$1$  &$1$  &$-2 $     &$-{2}/{7}$ &${5}/{7}$ &$5$\\
\hline
\hline
$\pq$$_3$&1&1&$\sb$&$\sbb$&$\sb$&$-{1}/{3}$&$-{1}/{21}$&$ {2}/{7}$&$5$\\
$\pl$$_3$&1&1&$\sb$& $1$  &$\sb$&$1$       &$ {2}/{7} $&$-{5}/{7}$&$5$\\
\hline
$\pu$$_3$&1&1&$\sbb$&$\sb$&$1$  &$ {4}/{3}$&${1}/{21}$ &$-{2}/{7}$&$4$ \\
$\pd$$_3$&1&1&$\sbb$&$\sb$&$1$  &$-{2}/{3}$&${1}/{21}$ &$-{2}/{7}$&$4$ \\
$\pn$$_3$&1&1&$\sbb$&$1$  &$1$  &$ 0 $     &$-{2}/{7}$ &${5}/{7}$ &$4$\\
$\pe$$_3$&1&1&$\sbb$&$1$  &$1$  &$-2 $     &$-{2}/{7}$ &${5}/{7}$ &$4$\\
\hline
\end{tabular}
\caption {Preon and anti-preon fields in Model {\it ii})}
\tblab{3pre2}
\end{center}
\end{table}
\begin{table}[h]
\begin{center}
\begin{tabular}{|c||c|c|c||c|c|c||c|}
\hline
      & $SU(3)_C$ & $SU(2)_L$ & $U(1)_Y$ 
      &$U(1)_B$ & $U(1)_L$ & {\bf Z}$_8$ & baryon$(\times 1 / \Lambda^6)$ \\
\hline
\hline
$Q_1$&$\sb$&$\sb$&${1}/{3}$&${1}/{3}$&$0$&7&$\pq^5$$_1$ $\pl^2$$_1$ \\
$l_1$&$1$  &$\sb$&$-1$     &$ 0 $    &$1$&7&$\pq^6$$_1$ $\pl$$_1$   \\
\hline
$\bar{U}_1$&$\sbb$&$1$&$-{4}/{3}$&$-{1}/{3}$&$0$&2&
                 $\pu^2$$_1$$\pd^3$$_1$$\pe$$_1$$\pn$$_1$ \\
$\bar{D}_1$&$\sbb$&$1$&$ {2}/{3}$&$-{1}/{3}$&$0$&2&
                 $\pu^3$$_1$$\pd^2$$_1$$\pe$$_1$$\pn$$_1$ \\
$\bar{N}_1$&$1$   &$1$&$ 0 $     &$ 0 $     &$-1$&2&
                 $\pu^3$$_1$$\pd^3$$_1$$\pe$$_1$ \\
$\bar{E}_1$&$1$   &$1$&$ 2 $     &$ 0 $     &$-1$&2&
                 $\pu^3$$_1$$\pd^3$$_1$$\pn$$_1$ \\
\hline
\hline
$Q_2$&$\sb$&$\sb$&${1}/{3}$&${1}/{3}$&$0$&5&$\pq^5$$_2$ $\pl^2$$_2$ \\
$l_2$&$1$  &$\sb$&$-1$     &$ 0 $    &$1$&5&$\pq^6$$_2$ $\pl$$_2$   \\
\hline
$\bar{U}_2$&$\sbb$&$1$&$-{4}/{3}$&$-{1}/{3}$&$0$&3&
                 $\pu^2$$_2$$\pd^3$$_2$$\pe$$_2$$\pn$$_2$ \\
$\bar{D}_2$&$\sbb$&$1$&$ {2}/{3}$&$-{1}/{3}$&$0$&3&
                 $\pu^3$$_2$$\pd^2$$_2$$\pe$$_2$$\pn$$_2$ \\           
$\bar{N}_2$&$1$   &$1$&$ 0 $     &$ 0 $     &$-1$&3&
                 $\pu^3$$_2$$\pd^3$$_2$$\pe$$_2$ \\
$\bar{E}_2$&$1$   &$1$&$ 2 $     &$ 0 $     &$-1$&3&
                 $\pu^3$$_2$$\pd^3$$_2$$\pn$$_2$ \\
\hline
\hline
$Q_3$&$\sb$&$\sb$&${1}/{3}$&${1}/{3}$&$0$&3&$\pq^5$$_3$ $\pl^2$$_3$ \\
$l_3$&$1$  &$\sb$&$-1$     &$ 0 $    &$1$&3&$\pq^6$$_3$ $\pl$$_3$   \\
\hline
$\bar{U}_3$&$\sbb$&$1$&$-{4}/{3}$&$-{1}/{3}$&$0$&4&
                 $\pu^2$$_3$$\pd^3$$_3$$\pe$$_3$$\pn$$_3$ \\
$\bar{D}_3$&$\sbb$&$1$&$ {2}/{3}$&$-{1}/{3}$&$0$&4&
                 $\pu^3$$_3$$\pd^2$$_3$$\pe$$_3$$\pn$$_3$ \\
$\bar{N}_3$&$1$   &$1$&$ 0 $     &$ 0 $     &$-1$&4&
                 $\pu^3$$_3$$\pd^3$$_3$$\pe$$_3$ \\
$\bar{E}_3$&$1$   &$1$&$ 2 $     &$ 0 $     &$-1$&4&
                 $\pu^3$$_3$$\pd^3$$_3$$\pn$$_3$ \\
\hline
\end{tabular}
\caption { ``Baryon'' and ``anti-baryon'' fields in Model {\it ii})}
\tblab{3BB2}
\end{center}
\end{table}
\begin{table}[h]
\begin{center}
\begin{tabular}{|c||c|c|c||c|c|c||c|}
\hline
      & $SU(3)_C$ & $SU(2)_L$ & $U(1)_Y$ 
      &$U(1)_B$ & $U(1)_L$ & {\bf Z}$_8$ & meson$(\times /\Lambda)$\\
\hline
\hline
$\Phiu$$_1$  &$1$     &$\sb$&   $ 1$   &    $0$   & $0$&7&$\pq$$_1$$\pu$$_1$\\
$\Phid$$_1$  &$1$     &$\sb$&   $-1$   &    $0$   & $0$&7&$\pq$$_1$$\pd$$_1$ \\
$\Gu$$_1$    &{\bf Ad}&$\sb$&   $ 1$   &    $0$   & $0$&7&$\pq$$_1$$\pu$$_1$ \\
$\Gd$$_1$    &{\bf Ad}&$\sb$&   $-1$   &    $0$   & $0$&7&$\pq$$_1$$\pd$$_1$ \\
$X$$_1$      &$\sbb$  &$\sb$&$-{1}/{3}$&$-{1}/{3}$& $1$&7&$\pq$$_1$$\pn$$_1$ \\
$\bar{X}$$_1$&$\sb$   &$\sb$&$ {1}/{3}$&$ {1}/{3}$&$-1$&7&$\pl$$_1$$\pd$$_1$ \\
$Y$$_1$      &$\sbb$  &$\sb$&$-{7}/{3}$&$-{1}/{3}$& $1$&7&$\pq$$_1$$\pe$$_1$ \\
$\bar{Y}$$_1$&$\sb$   &$\sb$&$ {7}/{3}$&$ {1}/{3}$&$-1$&7&$\pl$$_1$$\pu$$_1$ \\
$\phiu$$_1$  &$1$     &$\sb$&   $ 1$   &    $0$   & $0$&7&$\pl$$_1$$\pn$$_1$ \\
$\phid$$_1$  &$1$     &$\sb$&   $-1$   &    $0$   & $0$&7&$\pl$$_1$$\pe$$_1$ \\
\hline
\hline
$\Phiu$$_2$  &$1$     &$\sb$&   $ 1$   &    $0$   & $0$&0&$\pq$$_2$$\pu$$_2$\\
$\Phid$$_2$  &$1$     &$\sb$&   $-1$   &    $0$   & $0$&0&$\pq$$_2$$\pd$$_2$ \\
$\Gu$$_2$    &{\bf Ad}&$\sb$&   $ 1$   &    $0$   & $0$&0&$\pq$$_2$$\pu$$_2$ \\
$\Gd$$_2$    &{\bf Ad}&$\sb$&   $-1$   &    $0$   & $0$&0&$\pq$$_2$$\pd$$_2$ \\
$X$$_2$      &$\sbb$  &$\sb$&$-{1}/{3}$&$-{1}/{3}$& $1$&0&$\pq$$_2$$\pn$$_2$ \\
$\bar{X}$$_2$&$\sb$   &$\sb$&$ {1}/{3}$&$ {1}/{3}$&$-1$&0&$\pl$$_2$$\pd$$_2$ \\
$Y$$_2$      &$\sbb$  &$\sb$&$-{7}/{3}$&$-{1}/{3}$& $1$&0&$\pq$$_2$$\pe$$_2$ \\
$\bar{Y}$$_2$&$\sb$   &$\sb$&$ {7}/{3}$&$ {1}/{3}$&$-1$&0&$\pl$$_2$$\pu$$_2$ \\
$\phiu$$_2$  &$1$     &$\sb$&   $ 1$   &    $0$   & $0$&0&$\pl$$_2$$\pn$$_2$ \\
$\phid$$_2$  &$1$     &$\sb$&   $-1$   &    $0$   & $0$&0&$\pl$$_2$$\pe$$_2$ \\
\hline
\hline
$\Phiu$$_3$  &$1$     &$\sb$&   $ 1$   &    $0$   & $0$&1&$\pq$$_3$$\pu$$_3$\\
$\Phid$$_3$  &$1$     &$\sb$&   $-1$   &    $0$   & $0$&1&$\pq$$_3$$\pd$$_3$ \\
$\Gu$$_3$    &{\bf Ad}&$\sb$&   $ 1$   &    $0$   & $0$&1&$\pq$$_3$$\pu$$_3$ \\
$\Gd$$_3$    &{\bf Ad}&$\sb$&   $-1$   &    $0$   & $0$&1&$\pq$$_3$$\pd$$_3$ \\
$X$$_3$      &$\sbb$  &$\sb$&$-{1}/{3}$&$-{1}/{3}$& $1$&1&$\pq$$_3$$\pn$$_3$ \\
$\bar{X}$$_3$&$\sb$   &$\sb$&$ {1}/{3}$&$ {1}/{3}$&$-1$&1&$\pl$$_3$$\pd$$_3$ \\
$Y$$_3$      &$\sbb$  &$\sb$&$-{7}/{3}$&$-{1}/{3}$& $1$&1&$\pq$$_3$$\pe$$_3$ \\
$\bar{Y}$$_3$&$\sb$   &$\sb$&$ {7}/{3}$&$ {1}/{3}$&$-1$&1&$\pl$$_3$$\pu$$_3$ \\
$\phiu$$_3$  &$1$     &$\sb$&   $ 1$   &    $0$   & $0$&1&$\pl$$_3$$\pn$$_3$ \\
$\phid$$_3$  &$1$     &$\sb$&   $-1$   &    $0$   & $0$&1&$\pl$$_3$$\pe$$_3$ \\\hline
\end{tabular}
\caption { ``Meson'' fields in Model {\it ii})}
\tblab{3M2}
\end{center}
\end{table}
\begin{table}[h]
\begin{center}
\begin{tabular}{|c||c|c|c|c||c|c|}
\hline
      & $SU(7)_{H}$ 
      & $SU(3)_C$ & $SU(2)_L$ & $U(1)_Y$ 
      & $U(1)_B$  & $U(1)_L$ \\
\hline
\hline
$\pq$$_1$ & $1$   & $\sb$ & $\sb$ & $1/3$ & $1/3$ & $ 0$  \\
$\pl$$_1$ & $1$   &  $1$  & $\sb$ &  $1$  &  $0$  & $-1$  \\
\hline
$\pu$$_1$ & $1$   &$\sbb$ & $1$  & $-4/3$ & $-1/3$    & $0$ \\
$\pd$$_1$ & $1$   &$\sbb$ & $1$  & $ 2/3$ & $-1/3$    & $0$ \\
$\pn$$_1$ & $1$   & $1$   & $1$  & $ 0  $ & $ 0  $    & $-1$ \\
$\pe$$_1$ & $1$   & $1$   & $1$  & $ 2 $  & $ 0  $    & $-1$ \\
\hline
\hline
$\pq$$_2$ & $1$   & $\sb$ & $\sb$ & $1/3$ & $1/3$ & $ 0$  \\
$\pl$$_2$ & $1$   &  $1$  & $\sb$ &  $1$  &  $0$  & $-1$  \\
\hline
$\pu$$_2$ & $1$   &$\sbb$ & $1$  & $-4/3$ & $-1/3$    & $0$ \\
$\pd$$_2$ & $1$   &$\sbb$ & $1$  & $ 2/3$ & $-1/3$    & $0$ \\
$\pn$$_2$ & $1$   & $1$   & $1$  & $ 0  $ & $ 0  $    & $-1$ \\
$\pe$$_2$ & $1$   & $1$   & $1$  & $ 2 $  & $ 0  $    & $-1$ \\
\hline
\hline
$\pq$$_3$ &$\sb$  & $\sbb$& $\sb$ & $-1/3$&$-{1}/{21}$&$ {2}/{7}$ \\
$\pl$$_3$ &$\sb$  & $1$   & $\sb$ & $1$   &$ {2}/{7} $&$-{5}/{7}$ \\
\hline
$\pu$$_3$ &$\sbb$ & $\sb$ & $1$  & $ 4/3$ & $1/21$    & $-2/7$ \\
$\pd$$_3$ &$\sbb$ & $\sb$ & $1$  & $-2/3$ & $1/21$    & $-2/7$ \\
$\pn$$_3$ &$\sbb$ & $1$   & $1$  & $ 0 $  & $-2/7$    & $ 5/7$ \\
$\pe$$_3$ &$\sbb$ & $1$   & $1$  & $-2 $  & $-2/7$    & $ 5/7$ \\
\hline
\end{tabular}
\caption {Preon and anti-preon fields in Model {\it iii})}
\tblab{3pre3}
\end{center}
\end{table}
\begin{table}[h]
\begin{center}
\begin{tabular}{|c||c|c|c|c||c|c||c|}
\hline
      & $SU(7)_H$
      & $SU(3)_C$ & $SU(2)_L$ & $U(1)_Y$ 
      &$U(1)_B$ & $U(1)_L$    &  baryon $(\times 1/\Lambda^6)$\\
\hline
\hline
$Q_3$ &1& $\sb$&$\sb$&${1}/{3}$&${1}/{3}$&$0$&$\pq^5$$_3$ $\pl^2$$_3$ \\
$l_3$ &1& $1$  &$\sb$&$-1$     &$ 0 $    &$1$&$\pq^6$$_3$ $\pl$$_3$   \\
\hline
\hline
$\bar{U}_3$&1&$\sbb$&$1$&$-{4}/{3}$&$-{1}/{3}$&$0$&
                 $\pu^2$$_3$$\pd^3$$_3$$\pe$$_3$$\pn$$_3$ \\
$\bar{D}_3$&1&$\sbb$&$1$&$ {2}/{3}$&$-{1}/{3}$&$0$&
                 $\pu^3$$_3$$\pd^2$$_3$$\pe$$_3$$\pn$$_3$ \\
$\bar{N}_3$&1&$1$   &$1$&$ 0 $     &$ 0 $     &$-1$&
                 $\pu^3$$_3$$\pd^3$$_3$$\pe$$_3$ \\
$\bar{E}_3$&1&$1$   &$1$&$ 2 $     &$ 0 $     &$-1$&
                 $\pu^3$$_3$$\pd^3$$_3$$\pn$$_3$ \\
\hline
\end{tabular}
\caption { ``Baryon'' and ``anti-baryon'' fields in Model {\it iii})}
\tblab{3BB3}
\end{center}
\end{table}
\begin{table}[h]
\begin{center}
\begin{tabular}{|c||c|c|c|c||c|c||c|}
\hline
      & $SU(7)_{H_1}$
      & $SU(3)_C$ & $SU(2)_L$ & $U(1)_Y$ 
      &$U(1)_B$ & $U(1)_L$ & meson $(\times 1/\Lambda)$\\
\hline
\hline
$\Phiu$$_3$&1&$1$     &$\sb$&   $ 1$   &    $0$   & $0$&$\pq$$_3$$\pu$$_3$\\
$\Phid$$_3$&1&$1$     &$\sb$&   $-1$   &    $0$   & $0$&$\pq$$_3$$\pd$$_3$ \\
$\Gu$$_3$  &1&{\bf Ad}&$\sb$&   $ 1$   &    $0$   & $0$&$\pq$$_3$$\pu$$_3$ \\
$\Gd$$_3$  &1&{\bf Ad}&$\sb$&   $-1$   &    $0$   & $0$&$\pq$$_3$$\pd$$_3$ \\
$X$$_3$    &1&$\sbb$  &$\sb$&$-{1}/{3}$&$-{1}/{3}$& $1$&$\pq$$_3$$\pn$$_3$ \\
$\bar{X}$$_3$&1&$\sb$   &$\sb$&$ {1}/{3}$&$ {1}/{3}$&$-1$&$\pl$$_3$$\pd$$_3$ \\
$Y$$_3$      &1&$\sbb$  &$\sb$&$-{7}/{3}$&$-{1}/{3}$& $1$&$\pq$$_3$$\pe$$_3$ \\
$\bar{Y}$$_3$&1&$\sb$   &$\sb$&$ {7}/{3}$&$ {1}/{3}$&$-1$&$\pl$$_3$$\pu$$_3$ \\
$\phiu$$_3$  &1&$1$     &$\sb$&   $ 1$   &    $0$   & $0$&$\pl$$_3$$\pn$$_3$ \\
$\phid$$_3$  &1&$1$     &$\sb$&   $-1$   &    $0$   & $0$&$\pl$$_3$$\pe$$_3$ \\
\hline
\end{tabular}
\caption {``Meson'' fields in Model {\it iii})}
\tblab{3M3}
\end{center}
\end{table}
\efoot
\end{document}